\documentclass[fleqn,usenatbib]{mnras}

\usepackage{newtxtext,newtxmath}

\usepackage[T1]{fontenc}
\usepackage{ae,aecompl}

\usepackage{graphicx}	
\usepackage{amsmath}	
\usepackage{amssymb}	

\newcommand{\hmxb}{HMXB}
\newcommand{\ls}{LS~5039}
\newcommand{\lsi}{LS~I~+61$^{\circ}$~303}
\newcommand{\hj}{HESS~J0632+057}
\newcommand{\fgl}{1FGL~J1018.6-5856}
\newcommand{\psr}{PSR~B1259-63}
\newcommand{\sm}{{\rm M}_{\odot}}

\newcommand{\ms}{M_*}
\newcommand{\vecr}[1]{\vec{r}_{#1}}
\newcommand{\rell}{\vec{r}_{\text{ell}}}
\newcommand{\co}{\cos \omega}
\newcommand{\so}{\sin \omega}
\newcommand{\cO}{\cos \Omega}
\newcommand{\sO}{\sin \Omega}

\newcommand{\vtho}{\overrightarrow{\theta_0}}
\newcommand{\vth}{\vec{\theta}}
\newcommand{\vx}[1]{\overrightarrow{X}_{#1}}
\newcommand{\veps}[1]{\vec{\epsilon}_{#1}}
\newcommand{\pf}[2]{\frac{\partial #1}{\partial #2}}

\newcommand{\mcmeas}{M_{\rm c,meas}}
\newcommand{\sigm}{\sigma_{\rm m}}
\newcommand{\sigmis}{\sigma_{\rm mis}}

\newcommand{\sj}{Small-JASMINE}

\title[Uncovering the identities of compact objects
]{Uncovering the identities of compact objects
in high-mass X-ray binaries and gamma-ray binaries by astrometric
measurements}

\author[M. S. Yamaguchi et al.]{
Masaki S. Yamaguchi,$^{1}$\thanks{E-mail: masaki@ioa.s.u-tokyo.ac.jp}
T. Yano,$^{2}$
and N. Gouda$^{2,3}$
\\
$^{1}$Institute of Astronomy, Graduate School of Science,
    University of Tokyo, Mitaka, Tokyo 181-0015, Japan\\
$^{2}$National Astronomical Observatory, Mitaka, Tokyo 181-8588,
     Japan\\
$^{3}$SOKENDAI (The Graduate University for Advanced Students),
Shonan Village, Hayama, Kanagawa 240-0193, Japan
}

\date{Accepted XXX. Received YYY; in original form ZZZ}

\pubyear{2017}

\begin{document}
\label{firstpage}
\pagerange{\pageref{firstpage}--\pageref{lastpage}}
\maketitle

\begin{abstract}
We develop a method for identifying a compact object in binary systems with astrometric measurements and apply it to some binaries.
Compact objects in some high-mass X-ray binaries and gamma-ray binaries are unknown, which is responsible for the fact that emission mechanisms in such systems have not yet confirmed. The accurate estimate of the mass of the compact object allows us to identify the compact object in such systems. Astrometric measurements are expected to enable us to estimate the masses of the compact objects in the binary systems via a determination of a binary orbit.
We aim to evaluate the possibility of the identification of the compact objects for some binary systems.
We then calculate probabilities that the compact object is correctly identified with astrometric observation (= confidence level) by taking into account a dependence of the orbital shape on orbital parameters and distributions of masses of white dwarfs, neutron stars, and
black holes.
We find that the astrometric measurements with the precision of 70$\mu$as for $\gamma$~Cas
allows us to identify the compact object at 99\% confidence level if the compact object is
a white dwarf with 0.6 $\sm$. In addition, we can identify the compact object with the precision of 10 $\mu$as at 97\% or larger confidence level for \lsi\ and 99\% or larger for \hj. These results imply that the astrometric measurements with the 10-$\mu$as precision level can realize the identification of compact objects for $\gamma$~Cas, \lsi, and \hj.
\end{abstract}

\begin{keywords}
astrometry -- high-mass X-ray binaries -- gamma-ray binaries
\end{keywords}



\section{Introduction}
\label{intro}

The binary emitting X-ray is one of the most important targets in
high-energy astrophysics.
This kind of binary is generally called X-ray binary, which consists of
an optical star and a compact object (a neutron star or a black hole).
Observations in X-ray bands allowed us to develop the theory of
the accretion disk,
and many black hole candidates have been discovered from them.
Recently, a new binary class called gamma-ray binary, from which gamma
rays are observed as well as X-ray (mentioned in detail below),
has been discovered \citep{aha05}.
The X-ray emissions from Gamma-ray binaries show power-law spectra,
which means that non-thermal electrons (and/or positrons) are accelerated
in the binary systems.
This implies that the emissions from gamma-ray binaries can provide a clue
to the solution of the long-standing problem of the particle acceleration.
Thus, these binaries become more and more important for the study
on high-energy astrophysics.

Emission mechanism has been under discussion for some kinds of binaries,
such as $\gamma$~Cas.
The binary $\gamma$~Cas is one of the peculiar Be/X-ray binaries,
which generally consist of a pulsar and a Be star
(B star with a circumstellar disk).
Although the X-ray source is associated with the Be star,
spectral features and X-ray variability of $\gamma$~Cas are
different from those of other typical Be/X-ray binaries.
The observational features of $\gamma$~Cas different from other Be/X-ray
binaries are thermal X-ray with high temperature (10--12 keV),
relatively low luminosity ($L_{\rm x} \sim 10^{32-33}$ erg/s),
and no flare activity \citep[][and reference therein]{lop10}.
Such peculiar features are observed in some other Be/X-ray binaries,
which are called $\gamma$~Cas like objects \citep{lop06}.
Two viable scenarios explaining these X-ray features of $\gamma$~Cas
are proposed;
(1) accretion of the matter from the circumstellar disk onto
a white dwarf \citep{lop07} or a neutron star \citep{whi82} and
(2) magnetic interaction between the stellar surface and
the circumstellar disk \citep[e.g.,][]{smi99}.
The compact object has not been identified in the former scenario,
so that it is unknown that the accretion in the system is similar
to that in a cataclysmic variable (a white dwarf) or an X-ray
pulsar (a neutron star).

Emission mechanisms in gamma-ray binary systems also have not
been understood, mainly because compact objects
in most of gamma-ray binaries have never been identified.
The gamma-ray binary is a new class of high-mass X-ray binaries (\hmxb),
which was first discovered in 2004
\citep{aha05}, and six objects have been identified so far.
Non-thermal radio to TeV gamma-ray emissions have been detected
from gamma-ray binaries, and some of these emissions vary with the
orbital period.
However, the emission mechanism has not yet been explained clearly
\citep[e.g.,][]{dub13}.
If the compact object is a neutron star, we expect that
the non-thermal emissions are produced at shocks formed by the
collision between the pulsar wind and the stellar wind.
If it is a black hole instead, we expect that the stellar wind
accretes to drive the jet, so that non-thermal emissions can be
produced at shocks in the jet (internal shock model) and/or shocks
between the jet and the stellar wind (external shock model).
Thus, the identification of the compact object allows us to constrain
the emission mechanism.

We can identify such unrevealed compact objects by measuring their
masses.
The masses of white dwarfs, neutron stars, and black holes show different
distributions. 
The mass distribution of the white dwarfs has a peak at 0.5--0.6 $\sm$ and
the masses for almost all of the white dwarfs are below $\sim$1 $\sm$ 
\citep{kep16}.
The masses of the neutron stars distribute between 1.0 and 2.0 $\sm$,
which are estimated
by measuring the mass of neutron stars in neutron star-white dwarf systems
and double neutron star systems \citep{kiz13}.
The mass range of black holes is 3--10 $\sm$, although their
masses have larger uncertainties \citep{sha09}.
Thus, we can statistically identify a compact object by determining
the mass of the compact object in a X-ray binary where the compact object
has not yet been identified.

Astrometric observations allow us to dynamically measure the mass of
such an unseen compact object.
The astrometric time-series data of the star with an unseen companion
makes it possible to determine all orbital
elements, by the use of a method in literature
\citep[e.g.,][]{mon78,asa04,asa08}.
Given the mass of the observed star in addition to its orbital elements
(especially, the semi-major axis and orbital period),
we can estimate the mass of the companion star by using the Kepler's
third law. 
Thus, it is possible that the astrometric measurements for X-ray binaries
and gamma-ray binaries constrain the mass of their compact objects
and determine the identity of them (white dwarfs, neutron stars, or
black holes).
This means that if a compact object in such binaries
is unknown, the compact object can be identified by performing
the high-precision astrometry.

Recent/future astrometric satellites, such as Gaia and Small-JASMINE,
can perform the observation for stellar positions and motions
with high precision.
The Gaia satellite was launched on 19 December 2013 and its mission period
is planned to be around five years. It surveys the whole sky and observes
a billion of stars
at G-band \citep[0.3--1.0 $\mu$m;][]{deb12}.
Gaia measures the positions of photocentres of stellar images
with a precision of
7 microarcsecond ($\mu$as) for objects brighter than 12 magnitude at G-band,
and it observes the same target once per $\sim$ 50 days.
Small-JASMINE is planned as the second mission of JASMINE project series
\citep[other two missions are called Nano-JASMINE and (medium-sized)
JASMINE;][]{gou11}.
It will survey the Galactic centre region and observe other some regions
including interesting scientific targets at Hw-band (1.1--1.7 $\mu$m).
Small-JASMINE will perform the astrometric observation with a precision of
$\sim$10 $\mu$as for objects brighter than 12.5 magnitude at Hw-band.
Small-JASMINE can observe a certain target once per 100 minutes,
which is much more frequent than the observational frequency of Gaia,
so that Small-JASMINE can detect short-term variabilities of the
photocentres of stellar motions.

The capability of Gaia to detect an unidentified compact companion 
(neutron star) in \hmxb s have been investigated
by \citet{mao05,mao06}.
The authors calculate the probability distribution function of
a semi-major axis of a target OB star which has an unseen companion
taking into account the mass transfer and supernova explosion.
They then find the probability that
the semi-major axis is within the range in which Gaia can
detect a neutron star as a companion of OB stars.
They obtain the probabilities for some known OB stars as a function of
the kick velocity due to the supernova explosion.
They also apply their ideas to known \hmxb s to obtain the results
that an unidentified companion in the $\gamma$ Cas system can be detected
with a probability of 30--50 \%, and for the X Per system
Gaia can detect a neutron star with a probability of 10--30 \%,
where the range of
these probabilities reflects the range of the kick velocity
(0--200 km/s).
In addition, by using the semi-major axes
for these two systems, which are estimated based on the data of
spectroscopic observations, they argue that the companions of
these binary systems can be detected by Gaia.
We should note that they discuss the possibility that Gaia can detect
a compact companion based only on the semi-major axis. 

The detectability of an unseen companion with astromeric observations
depends not only on the semi-major axis but also the other orbital
elements.
\citet{asa07} have investigated precisions of determining
orbital parameters by astrometric observations
for elliptical orbits with various values of the eccentricity,
inclination angle, and argument of periapsis.
They have shown that even if they do not change the semi-major axis,
the precision of semi-major axis changes by one order of magnitude.
The detection of semi-major axis of the target star means the detection
of an unseen companion, so that the precision of semi-major axis
can be a criterion of the detection of an unseen companion.
Thus, the results of \citet{asa07} implies that the eccentricity,
inclination angle, and argument of periapsis of the stellar orbit
affects the detectability of an unseen companion.
Therefore, we should take into account the orbital elements other
than the semi-major axis to estimate the possibility that a companion
can be detected by astrometric observations.

In this paper, we evaluate possibility that compact objects in specific \hmxb s and gamma-ray binaries 
are correctly identified, which corresponds to a confidence level.
To identify compact objects from their masses, we should
incorporate the distributions of white dwarfs, neutron stars,
and black holes into a procedure to estimate the astrometric precision.
In addition, we take into account these orbital parameters.
In this paper, we focus on nearby \hmxb s and gamma-ray binaries.
We demonstrate the way to calculate confidence levels
in Section \ref{sec:method} and
show the astrometric precisions required for the
identification of unknown compact objects for specific binary systems
in Section \ref{sec:results}.
We then discuss 
a feasibility of some astrometric missions in Section \ref{sec:dis}.
Finally, we summarize our study in Section \ref{sec:sum}.

\section{Methods}
\label{sec:method}
 In this section, we show the method to calculate confidence levels
for identifying the compact object in \hmxb s and gamma-ray binaries.
We derive an expression of a precision of the compact object mass as a function of
the compact object mass itself, astrometric precision, and the orbital parameters
by using an equation of the error propagation.
By giving a probability distribution function and defining the mass boundary
between a white dwarf, a neutron star, and a black hole, we show how to calculate
a probability that the compact object is correctly identified, which corresponds to
the confidence level, for a mass of the compact object and an astrometric precision.

\subsection{Precision of the compact object mass}
In this subsection, we formulate an expression of a precision of the compact object mass.
Before we adopt an equation of the error propagation, we show the relation equation between
the compact object mass $M_{\rm c}$, stellar mass $M_*$, orbital period $P_{\rm orb}$, 
and semi-major axis of the companion, which is represented as a product of an angular semi-major
axis of the companion $a_*$ and distance $D$:
\begin{equation}
\frac{(M_{\rm c}+M_*)^2}{M_{\rm c}^3} = \frac{G}{4\pi^2}\frac{P_{\rm orb}^2}{(a_*D)^3},
\label{eq:kepler}
\end{equation}
where $G$ is the gravitational constant. 
This equation means that the identification of compact objects requires measurements of $M_*$, $P_{\rm orb}$, $a_*$, and $D$ with a sufficient accuracy. 
By using Equation (\ref{eq:kepler}) and an equation of the error propagation, we obtain the expression
of the precision of the compact object mass $\sigma_{\rm m}$:
\begin{equation}
\label{eq:errprop}
\begin{split}
\left( \frac{\sigma_{\rm m}}{M_{\rm c}} \right)^2 =& \left( \frac{3}{2} - \frac{M_{\rm c}}{M_{\rm c}+M_*} \right)^{-2} \\ & \times
\left[ \left( \frac{M_*}{M_{\rm c}+M_*} \right)^2 \frac{\sigma_{\rm M*}^2}{M_*^2}+ \frac{\sigma_P^2}{P_{\rm orb}^2} +
\frac{9}{4}\left( \frac{\sigma_a^2}{a_*^2} + \frac{\sigma_D^2}{D^2} \right) \right],
\end{split}
\end{equation}
where $\sigma_{\rm M*}$, $\sigma_P$, $\sigma_a$, and $\sigma_D$ are the standard errors
of the companion mass, orbital period, semi-major axis, and distance, respectively.
In deriving this equation, we assume that for all variables the ratios of the standard errors to
the variables themselves are smaller than 1, and the correlation between errors of
 these parameters can be neglected.
Here, we should note that the correlation between errors of the semi-major axis and distance
may be non-negligible if the orbital period is close to 1 year, because the orbital motion is
similar to the annual elliptical motion in that case.

We can further reduce Equation (\ref{eq:errprop}) by adopting some approximation.
Orbital periods of X-ray binaries and gamma-ray binaries is usually well-determined
compared to the stellar mass and distance, so that we can neglect
the term of the standard error of the orbital period in Equation (\ref{eq:errprop}).
In addition, the term $\frac{\sigma_D^2}{D^2}$ is expressed as $\frac{\sigma_\pi^2}{\pi_{\rm p}^2}$,
where $\pi_{\rm p}$ and $\sigma_\pi$ are the parallax and its standard error, respectively,
and $\sigma_a$ can be reduced to be $f(e,i,\omega)\sigma_{\rm mis}$ (see Appendix \ref{sec:sigma_a}),
where $f(e,i,\omega)$ and $\sigma_{\rm mis}$ are the function of eccentricity $e$,
inclination angle $i$, and argument of periapsis $\omega$ (Appendix \ref{sec:order},
and precisions reachable with a mission, respectively.
Thus, Equation (\ref{eq:errprop}) can be reduced as
\begin{equation}
\begin{split}
\left( \frac{\sigma_{\rm m}}{M_{\rm c}} \right)^2 =& \left( \frac{3}{2} - \frac{M_{\rm c}}{M_{\rm c}+M_*} \right)^{-2} \\ & \times
\left[ \left( \frac{M_*}{M_{\rm c}+M_*} \right)^2 \frac{\sigma_{\rm M*}^2}{M_*^2}+
\frac{9}{4}\sigma_{\rm mis}^2\left( \frac{f^2}{a_*^2} + \frac{1}{\pi_{\rm p}^2} \right) \right],
\end{split}
\label{eq:errprop_red}
\end{equation}
where we assume that the precision of parallax can be approximated to be that reachable
with a mission.
We here note that $a_*$ is estimated by using $M_{\rm c}$, $M_*$, $P_{\rm orb}$, and $\pi_{\rm p}$.
Thus, this equation means that we obtain the precision of the compact object mass if we give $M_{\rm c}$,
$\sigma_{\rm mis}$, and parameters inherent to each object ($M_*, \sigma_{\rm M*}, f(e,i,\omega),
P_{\rm orb}, \text{ and } \pi_{\rm p}$).

\subsection{Confidence level}
\label{sec:reqmass}
\begin{figure}
 \begin{center}
 \includegraphics[width=8cm,clip]{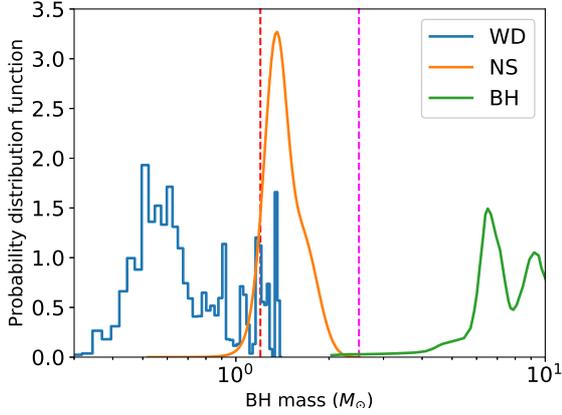}
 \caption{
	 Distributions of masses for white dwarfs, neutron stars, and black holes.
	 We also show the boundary mass between a white dwarf and a neutron star (1.2$\sm$; red line)
	 and one between a neutron star and a black hole (2.5$\sm$; magenta line),
	 which are adopted in this paper.
	 The mass distributions of white dwarfs, neutron stars, and black holes are
	 taken from those in Figure 3 of \citet{kep16}, Figure 2 of \citet{kiz13}, and
	 Figure 2 of \citet{oze10}, respectively.
	 }
 \label{fig:mass-dis}
 \end{center}
\end{figure}

In this subsection, we demonstrate the way to find confidence levels
by calculating the possibility that the compact object is correctly identified.
The boundary mass and the probability distribution function of measured mass
are required for calcuating this possibility.
We adopt boundary masses of 1.2 $\sm$ between a white dwarf and a neutron star
and 2.5 $\sm$ between a neutron star and a black hole (Figure \ref{fig:mass-dis}).
The boundary mass 1.2 $\sm$ is obtained so that a fraction of white dwarfs below the mass
is equal to that of neutron stars above the mass, i.e.,
\begin{equation}
\int^{M_{\rm bou}}_{0} F_{\rm WD}(M)dM = \int^{\infty}_{M_{\rm bou}} F_{\rm NS}(M)dM \sim 0.9,
\label{eq:mbou}
\end{equation}
where $M_{\rm bou}$ is the boundary mass between a white dwarf and a neutron star,
and $F_{\rm WD}$ and $F_{\rm NS}$ are the probability distribution functions for masses of
white dwarfs and neutron stars, respectively.
In addition, we should note that $F_{\rm NS}$ has two components; NS-WD system and double NS system.
Here, we assume that contributions to all neutron stars of these two components is equal.
The last equation of Equation (\ref{eq:mbou}) means that tenth of all white dwarfs has a mass
larger than 1.2 $\sm$ and tenth of all neutron stars has a mass smaller than 1.2 $\sm$.
The boundary mass between a neutron star and a black hole, 2.5 $\sm$, is determined to be a mass where
mass distributions of both neutron stars and black holes is nearly zero.
This boundary mass is also adopted in \citet{bel08}.
We use these boundary masses for judging the identity of the compact object.

We then obtain the confidence level by using probability distribution function
of measured mass.
Here, we assume that measured mass of a compact object $\mcmeas$ is normally distributed, i.e.,
\begin{equation}
N(\mcmeas; M_{\rm c},\sigm) \equiv \frac{1}{\sqrt{2\pi}\sigm}\exp \left(- \frac{(\mcmeas - M_{\rm c})^2}{2\sigm^2} \right),
\label{eq:pmmeas}
\end{equation}
where we note that $M_{\rm c}$ is a true value of a compact object mass.
We further assume that $M_{\rm c}$ distributes as in Figure \ref{fig:mass-dis},
so that we find the distribution of measured mass averaged using the distributions of $M_{\rm c}$:
\begin{equation}
\begin{split}
A_{\rm [ WD,NS,BH ]}(\mcmeas; \sigma_{\rm mis}) \equiv & \int N(\mcmeas; M_{\rm c},\sigm)
\\ & \ \ \times F_{\rm [ WD,NS,BH ]}(M_{\rm c}) dM_{\rm c},
\end{split}
\label{eq:ammeas}
\end{equation}
where $F_{\rm BH}$ is the distribution of black hole mass calculated in \citet{oze10},
and we note that the function $A$ also different from object to object.
In the case that the candidate of the compact object is a white dwarf or a neutron star,
the probability that we can correctly identify the compact object $P_{\rm [WD,NS]}$
can be expressed as follows:
\begin{align}
P_{\rm WD}(\mcmeas; \sigma_{\rm mis}) = \frac{A_{\rm WD}}{A_{\rm WD}+A_{\rm NS}} & \text{ for }
\mcmeas < 1.2 \sm,\label{eq:wn_wd}\\
P_{\rm NS}(\mcmeas; \sigma_{\rm mis}) = \frac{A_{\rm NS}}{A_{\rm WD}+A_{\rm NS}} & \text{ for }
\mcmeas > 1.2 \sm.\label{eq:wn_ns}
\end{align}
In the case that the candidate of the compact object is a neutron star or a black hole,
the probability that we can correctly identify the compact object $Q_{\rm [NS,BH]}$
can be expressed as follows:
\begin{align}
Q_{\rm NS}(\mcmeas; \sigma_{\rm mis}) = \frac{A_{\rm NS}}{A_{\rm NS}+A_{\rm BH}} & \text{ for }
\mcmeas < 2.5 \sm,\label{eq:nb_ns}\\
Q_{\rm BH}(\mcmeas; \sigma_{\rm mis}) = \frac{A_{\rm BH}}{A_{\rm NS}+A_{\rm BH}} & \text{ for }
\mcmeas > 2.5 \sm.\label{eq:nb_bh}
\end{align}
Thus, if $\mcmeas$ and $\sigma_{\rm mis}$ are given, we can calculate the confidence level
for each binary by using these expressions.
Here, we should note that prior probabilities of a white dwarf, a neutron star, and a black hole are identical, because the distributions shown in Figure \ref{fig:mass-dis} are all normalised as unity.

\section{Application to a HMXB and gamma-ray binaries}
\label{sec:results} 

\begin{table*}
	\centering
	\caption{A high-mass X-ray binaries selected from the catalogue,
``Catalogue of Galactic high-mass X-ray binaries" \citep{liu06} and
five gamma-ray binaries.
}
	\label{tab:hmxb}
	\begin{tabular}{lrrrrrrrl} 
		\hline
		 Object name &  V (mag) &  H (mag) &
 $M_*$ (M$_{\odot}$)  &  Parallax (mas) &
 Distance (kpc) &  P (days) &
 $f(e,i,\omega)$ &  Pulsar? \\
		\hline
  $\gamma$ Cas & 1.6$^a$ & 2.0$^a$ & 19$\pm$ 0.1$^b$
    & 5.9$^c$ & & 204$^a$ & 1.4 & unknown \\
  X Per        & 6.0$^a$ & 6.1$^a$ & 15$\pm$ 0.8$^b$
    & 2.3$^c$ & & 250$^a$ & 1.4 & yes\\
  V725 Tau     & 8.9$^a$ & 8.3$^a$ & 8$\pm$ 0.3$^b$
    & 3.7$^c$ & & 111$^a$ & 1.4 & yes \\
  V801 Cen     & 9.3$^a$ & 8.6$^a$ & 9$\pm$ 0.4$^b$
    & 1.8$^c$ & & 188$^a$ & $>$1.4 & yes \\
  \ls       & 11.2 & 8.8$^d$ & 23$\pm$ 3$^e$ & & 2.5$^f$ 
    & 3.9$^f$ & 1.4  & unknown \\
  \fgl   & 12.9 & 10.1$^d$& $\sim$ 37$^e$ & & 5$^f$ 
    & 16$^f$ & (1.4) & unknown \\
  \lsi   & 10.8 & 8.2$^d$ & 12.5$\pm$ 2.5$^e$ & & 1.9$^f$
    & 26$^f$ & 1.3  & unknown \\
  \hj     & 9.1  & 7.4$^d$ & 16$\pm$ 3$^e$ & & 1.4$^f$
    & 320$^f$ & 3.8-5.4 & unknown \\
  PSR B1259-63  & 9.3  & 8.6$^d$ & 31$^e$ & & 2.3$^f$
    & 1237$^f$ & 3.6-3.7 & yes \\
		\hline
	\end{tabular}
{\footnotesize \\ The former four objects are categorized as high-mass
X-ray binaries, and the others are gamma-ray binaries.\\
$^a$ Appeared in the HMXB catalogue \citep{liu06}.
$^b$ Appeared in the catalogue
   ``A catalog of young runaway {\it Hipparcos} stars within 3 kpc
   from the Sun" \citep{tet11}.
$^c$ Values appeared in {\it Hipparcos} catalogue
   \citep{van07}.
$^d$ 2MASS All-Sky Catalog of Point Sources \citep{skr06}.
$^e$ We adopt the middle values of the mass ranges appeared
   in \citet{cas12}.
$^f$ Values appeared in \citet{cas12}.
}
\end{table*}

We apply Equations (\ref{eq:wn_wd}-\ref{eq:nb_bh})
to a \hmxb\ and gamma-ray binaries.
We search for nearby \hmxb s with long orbital periods
because they are promising for the detection of
the orbit by astrometric observations.
Four good candidates are found from the HMXB catalogue \citep{liu06}
and listed up in Table \ref{tab:hmxb}.
Three objects other than $\gamma$ Cas are known to include
a neutron star,
so that we focus on $\gamma$ Cas in Section \ref{sec:gcas}.
Although five gamma-ray binaries have been detected in Milky Way,
the compact objects in the four binaries of them remain
unknown \citep[e.g.,][]{dub13}, so that we focus on these
four gamma-ray binaries in Section \ref{sec:grby}.

\begin{figure}
 \begin{center}
 \includegraphics[width=8cm,clip]{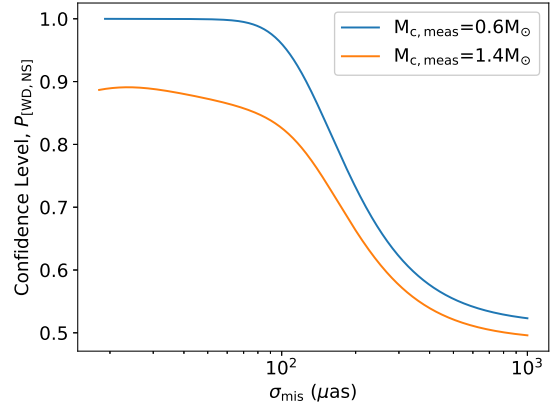}
 \caption{Confidence level as a function of the astrometric precision of a mission for $\gamma$ Cas. We show the cases that the measured masses of a compact object are 0.6 $\sm$ (blue) and 1.4 $\sm$ (orange).
	 }
 \label{fig:cl-gcas}
 \end{center}
\end{figure}

\subsection{$\gamma$ Cas (a white dwarf or a neutron star)}
\label{sec:gcas}
A white dwarf or a neutron star is acceptable as the compact object
in $\gamma$ Cas, given the dynamical mass derived with
the observation of radial velocity and an assumption of coplanarity
between the orbital plane and circumstellar disk \citep{har00}.
We calculate the precisions of $f(e,i,\omega)$ for $\gamma$ Cas
in the way stated in Appendix \ref{sec:cal}.
The eccentricity and argument of periapsis of $\gamma$ Cas are measured
as 0.26 and $\sim$50$^{\circ}$, respectively,
from the observation of H$\alpha$ emission \citep{har00},
so that $f(e, i, \omega)$ of $\gamma$ Cas is calculated as 1.4.
Here, we note that $f(e,i,\omega)$ is nearly independent of
the inclination angle when the argument of periapsis $\omega$ is small,
which is shown in the second paragraph of Appendix \ref{sec:calres}.
In addition, as shown in Table \ref{tab:hmxb}, the mass of companion star
in the $\gamma$ Cas system, the parallax, and the orbital period are
19 $\pm$0.1$\sm$, 5.9 mas, and 204 days, respectively.

We obtain the relations between confidence levels and $\sigmis$ for given measured masses,
which are shown in Figure \ref{fig:cl-gcas}. 
Here, we assume that the measured masses are 0.6 $\sm$ (a white dwarf is assumed and Equation \ref{eq:wn_wd} is used) and
1.4 $\sm$ (a neutron star is assumed and Equation \ref{eq:wn_ns} is used).
For a white dwarf case, the confidence level reaches $\simeq$1.0 below $\sim$70 $\mu$as,
while it is below 0.9 at a maximum for a neutron star case.
This is because the distribution of white dwarf mass (Figure \ref{fig:mass-dis}) extends to $\sim$1.4 $\sm$,
while that of neutron star mass shows nearly zero at 1.0 $\sm$.
Even if the astrometric precision is high enough, a white dwarf with $\sim$1.4 $\sm$ is expected to exist,
so that the probability that the compact object with a mass of 1.4 $\sm$ is a white dwarf is non-zero.
The minimum values for $\sigmis$ is determined by the corresponding $\sigm$ of 0.05 $\sm$,
because the intervals of bins in the distribution of white dwarf mass is $\sim$0.03, which means that
the integration of Equation (\ref{eq:ammeas}) is affected by one normal distribution corresponding to one bin
if $\sigm \lesssim$0.03.

Based on Figure \ref{fig:cl-gcas}, we argue that $\sigmis \sim$70  $\mu$as is enough to identify the compact object
in $\gamma$ Cas if the compact object has the typical mass of a white dwarf or a neutron star.
With such $\sigmis$, we can identify a white dwarf with a typical mass (0.5-0.6 $\sm$) at $\sim$99 \% confidence level,
where we can expect that in the case of a white dwarf with 0.5 $\sm$, the confidence level gets higher than that with 0.6 $\sm$,
because 0.5 $\sm$ is farther from the mass distribution of neutron stars.
In addition, we can identify a neutron star with a typical mass ($\sim$1.4 $\sm$) with $\sim$90 \% confidence level.


\begin{figure*}
 \begin{center}
 \includegraphics[width=8cm,clip]{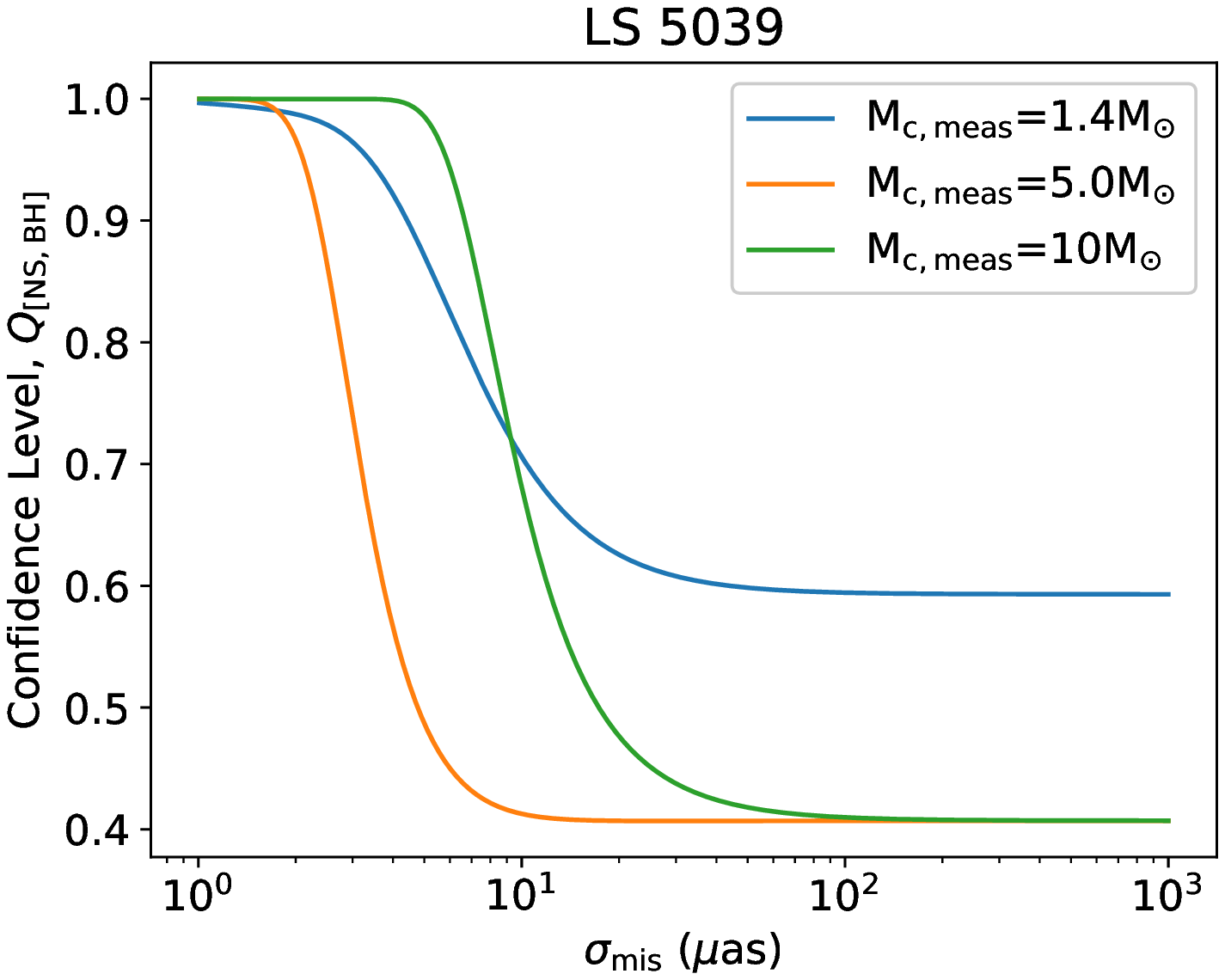}
 \includegraphics[width=8cm,clip]{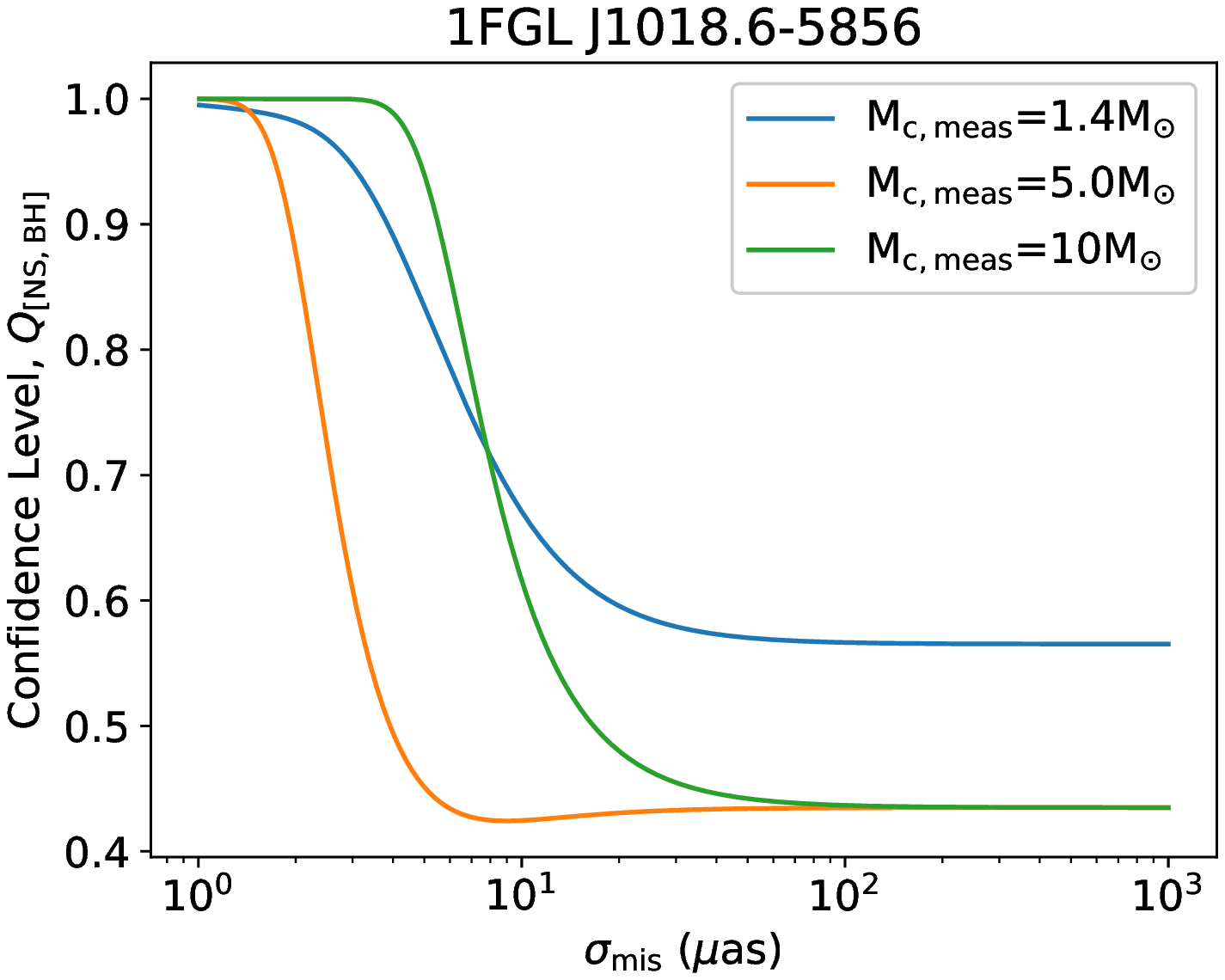}
 \includegraphics[width=8cm,clip]{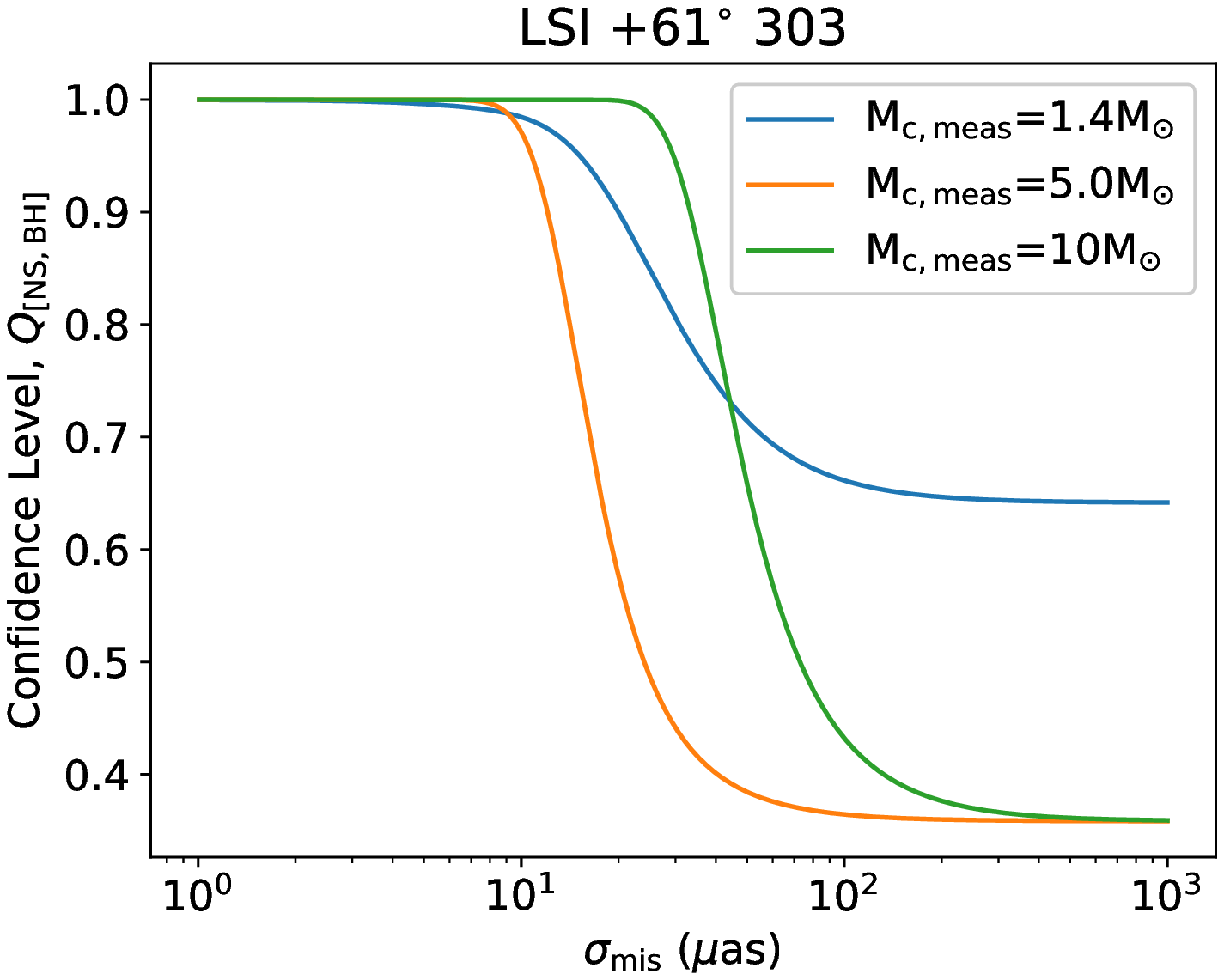}
 \includegraphics[width=8cm,clip]{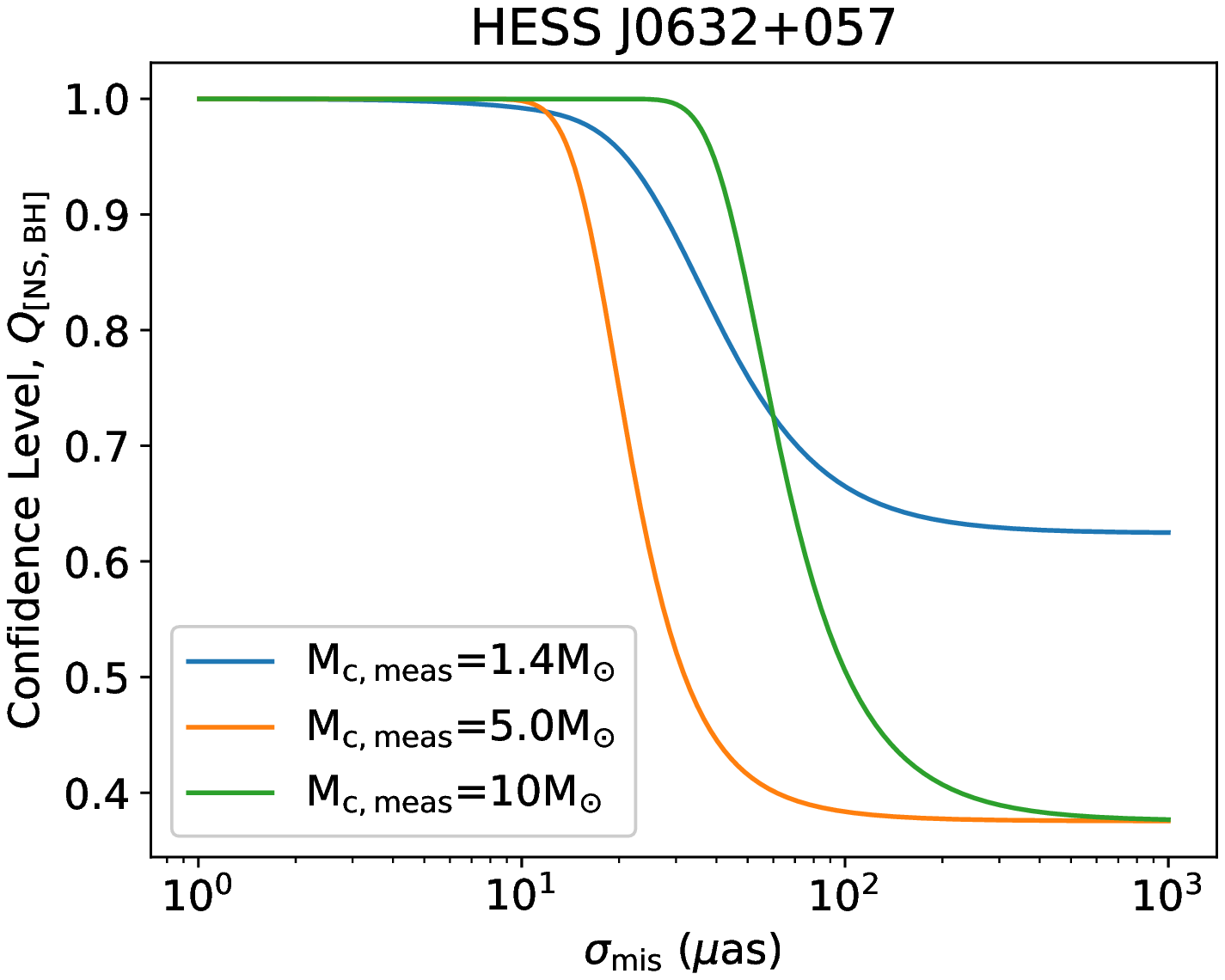}
 \caption{Same as Figure \ref{fig:cl-gcas} for gamma-ray binaries. We adopt the measured masses of 1.4, 5.0, and 10 $\sm$.
	 }
 \label{fig:cl-grby}
 \end{center}
\end{figure*}

\subsection{Gamma-ray binaries (a neutron star or a black hole)}
\label{sec:grby}
Applying Equations (\ref{eq:nb_ns} and \ref{eq:nb_bh}) to gamma-ray binaries allows us to
obtain confidence levels for determining whether the compact
object is a neutron star or a black hole.
In Table \ref{tab:hmxb}, we show the V and H magnitudes, the stellar masses,
the distances from the Earth, the orbital periods, and the functions
$f(e, i, \omega)$ for five galactic gamma-ray binaries.
Here, the error of stellar mass for \fgl\ is not shown, so that we adopt $\sigma_{\rm M*}=$5 $\sm$.
This is just an assumption, but the results do not change if $\sigma_{\rm M*}$ is less than the stellar mass itself,
because the dominant term in the right-hand side of Equation (\ref{eq:errprop_red}) is the term of $\sigmis$ for \fgl.

The functions $f(e,i,\omega)$ are calculated by using
known values of $e$ and $\omega$ 
We adopt $e=0.24$ and $\omega = 237^{\circ}$ for \ls\ \citep{sar11},
$e=0.54$ and $\omega = 41^{\circ}$ for \lsi\ \citep{ara09},
$e=0.83$ and $\omega = 129^{\circ}$ for \hj\ \citep{cas12}, and
$e=0.87$ and $\omega = 139^{\circ}$ for \psr\ \citep{joh94}.
For the inclination angle in the function $f$, we adopt 10$^{\circ}$
and 80$^{\circ}$ because of its uncertainty in the radial velocity measurements.
If the eccentricity is large and the argument of periapsis is close to
90$^{\circ}$, the uncertainty of the inclination angle affects
the $f(e,i,\omega)$ values (Appendix \ref{sec:calres}).
Thus, the $f(e,i,\omega)$ values of \hj\ and \psr\ are affected
by the uncertainty of $i$, so that Table \ref{tab:hmxb} shows
the ranges of the $f(e,i,\omega)$ values for \hj\ and \psr.
Here, we should note that the $f$ values reach 4 or 5 for
the high eccentric systems (\hj\ and \psr), which means that if we
neglect this factor, we overestimate the required astrometric
precision by a factor of 4 or 5 for high eccentric systems.
The orbital elements of \fgl\ have not yet determined,
so that we assume the typical value, 1.4.
Here, we adopt a middle value, 4.6, as $f$ of \hj.

We show the confidence levels for four gamma-ray binaries in Figure \ref{fig:cl-grby},
adopting the measured mass of 1.4, 5.0, and 10 $\sm$, where we assume that the compact
object masses of 5.0 and 10 $\sm$ are less and more massive black holes, respectively.
Confidence levels of all binaries at the minimum $\sigmis$ reach 1.0, which is because
the distributions of neutron star mass and black hole mass show little overlap (Figure \ref{fig:mass-dis}),
so that we can certainly identify the compact object for small $\sigmis$, that is, small $\sigm$.
Figure \ref{fig:cl-grby} also shows that the confidence levels of the measured mass 5.0 $\sm$ is higher than
those of 10 $\sm$. This is due to the difference between the measured mass and the mass range of the neutron star.
When the measured mass is farther from the mass range of neutron star, $A_{\rm NS}$ is smaller compared to
$A_{\rm BH}$, so that the confidence level is higher.

In addition, we see for all binaries that the confidence level of $\mcmeas=1.4\ \sm$ at the maximum $\sigmis$
is larger than 0.5, while that of $\mcmeas=5.0$ and 10 $\sm$ is smaller than 0.5.
This is due to the distributions of neutron star mass and black hole mass.
For a large $\sigm$, the distribution of measured mass can be approximated to be
$(2\pi)^{-1/2}\sigm^{-1}$ (see Equation \ref{eq:pmmeas}),  so that $Q_{\rm NS}$ and $Q_{\rm BH}$ are proportional to
$<\sigm^{-1}>_{\rm NS}$ and $<\sigm^{-1}>_{\rm BH}$, respectively, where $<x>_{[{\rm NS,BH}]}$
means the statistical average of $x$ for the distribution of neutron star mass or black hole mass.
We can derive $\sigm \sim M_{\rm c}\sigmis/a_*$ from Equation (\ref{eq:errprop_red})
for the current situation (large $\sigmis$, $a_*<\pi_{\rm p}$, and $M_{\rm c}< M_*$),
so that by using Equation (\ref{eq:kepler}), $\sigm \propto (M_{\rm c}+M_*)^{2/3}$.
This means that the larger $M_{\rm c}$ leads to the larger $\sigm$.
Therefore, $Q_{\rm NS}$ is larger than $Q_{\rm BH}$ at a large $\sigmis$.
This implies that if $\sigmis$ is large, a neutron star is tend to be chosen as a compact object,
even if the compact object is a black hole.

For \ls\ and \fgl, the astrometric precision of 2 $\mu$as is required for identifying the compact object
at high confidence level ($> 90\%$).
The confidence levels of these two binaries (upper panels in Figure \ref{fig:cl-grby}) are quite similar with each other,
because the values of $a_*$, which is a parameter related to the orbital size on the celestial sphere,
for two binaries are nearly identical. 
When we adopt 90\% as a required confidence level, the corresponding values of $\sigmis$ are $\sim$ 2 $\mu$as,
$\sim $4 $\mu$as, and $\sim $6 $\mu$as for $\mcmeas = 5.0, 1.4, \textrm{ and, } 10\ \ms$, respectively.
If we allow the confidence level to be 70\%, we can identify the compact object with $\sigmis = 10 \mu$as
for $\mcmeas = 1.4$ and 10 $\sm$.

For \lsi\ and \hj, the astrometric precision of 10 $\mu$as is enough for identifying the compact object.
When we adopt the 90\% confidence level, $\sigmis = 12, 20, \textrm{ and, }33 \mu$as are required for identifying
the compact object whose measured masses are 5.0, 1.4, and 10 $\sm$, respectively, for \lsi.
For \hj, the required astrometric precisions are 16, 28, and 44 $\mu$as for the measured masses, 5.0, 1.4, and
10 $\sm$, respectively.
In addition, when we adopt an astrometric precision of 10 $\mu$as, the confidence levels
for the measured mass of 1.4, 5.0, and 10 $\sm$
are 97\% or larger for \lsi\ and 99\% or larger for \hj.
These results implies that with 10-$\mu$as astromeric precision, we can identify the compact object in \lsi\ and \hj\ at >97\% confidence level if the measured mass is $\sim$1.4 $\sm$ or $>5.0 \sm$.

\section{Feasibility of the science objectives for the future/on-going
astrometric missions}
\label{sec:dis}
In this subsection, we discuss the feasibility of the science objectives
stated in Sections \ref{sec:results} 
by the future/on-going astrometric missions,
Gaia, \sj, GRAVITY of VLTI, and TMT.
All these missions aim to achieve the 10-$\mu$as level precision.
If the precision of 10 $\mu$as is achieved for the observation of binaries
listed in this paper, we will accomplish 
the identification of the compact object through the determination of
the compact object mass for $\gamma$ Cas, \lsi\ and \hj at $\sim$90\% or larger confidence level,
and for \ls\ and \fgl\ at 70\% confidence level.

The ground-based telescopes, TMT and GRAVITY, have a critical issue
in terms of reference stars for observing these objects.
The instrument IRIS imager will be used for astrometric observations with TMT.
To realise the 10-$\mu$as level astrometry with the IRIS imager,
some reference stars are required in a field of view of each
target object \citep{sch14}.
The field of view of the IRIS imager is 17" x 17",
so that it is necessary that some reference stars are located
within $\sim$ 10" from each target.
However, no source is found within 20" from $\gamma$~Cas,
\hj, and X Per according to the 2MASS catalogue \citep{skr06}.
In addition, although one source 5" away from V725 Tau is listed
in the 2MASS catalogue
\citep{skr06}, it is based on low quality results, and 
no other source is found within 20".
Therefore, the identification of the compact object for binaries listed in this paper
is not possible with TMT/IRIS.
The field of view of VLTI/GRAVITY is 2" for unit telescopes
or 4" for auxiliary telescopes \citep{eis11}, so that it is
thought to be impossible 
to realize the identification of the compact object.

The space telescopes for the astrometric observation, Gaia and
\sj, are promising for the identification of the compact
object. 
The fields of view of these satellites are the order of
1 square degree, where reference stars are included enough to
calibrate the stellar position.
Besides the field of view, to achieve these science objectives,
the observational
period longer than orbital periods of the binaries and
ten-time or more visits to the binaries are required
for measuring the position of whole orbits.
The Gaia satellite will visit to all sources on average
70 times for the five-year observation \citep{deb12}, and for \sj,
whose observational period is 3 years, we can
adjust the number of visits for individual source \citep{gou11}.
Therefore, these satellites will perform the 10-$\mu$as level astrometry 
without any problems in terms of not only reference stars
but also the observational period and the visit number.

\section{Conclusions}\label{sec:sum}

We have estimated in this paper probabilities that compact object in $\gamma$~Cas
and four gamma-ray binaries are correctly identified (= confidence levels) 
as a function of a compact object mass measured by astrometric observation of the binary orbits
and an astrometric precision.
In the procedure to estimate the confidence levels, 
we have incorporated the effect of the orbital elements 
(eccentricity, inclination angle, and argument of periapis) on
the precision of semi-major axis and the distributions
of masses of white dwarfs, neutron stars, and black holes, which are based on observations.
The former effect appears in the relation equation between
the precision of semi-major axis and the astrometric precision
(the function $f$ in Equation \ref{eq:osiga}), and the resultant
astrometric
precision is up to 5 times larger than that without this effect.
The later one is needed as a prior distribution of compact object masses.
We obtain three results as follows by applying this procedure
to \hmxb s and gamma-ray binaries.

(1) $\gamma$ Cas is one of the
nearest X-ray binaries and whether the compact object in this system
is a white dwarf or a neutron star remains unclear \citep[e.g.,][]{lop10}.
We have found that 
the astrometric measurements with the precision of 70 $\mu$as allows us to
identify the compact object as a white dwarf at 99\% confidence level if the measured mass is 0.6 $\sm$.
With such astrometric precision, we can identify the compact object as a neutron star at 85\% confidence level
if the measured mass is 1.4 $\sm$.

(2) Compact objects of four gamma-ray binaries, \ls, \fgl, \lsi, and
\hj, remains unknown.
For the former two systems, \ls\ and \fgl, the precision of 2 $\mu$as
is required for identifying the compact object as a black hole at 90\% confidence level
if the measured mass is 5.0 $\sm$.
If the measured mass is 1.4 or 10 $\sm$, we can identify the compact object with $\sim$10-$\mu$as
level astrometry at 70\% confidence level.

(3) For latter two systems, \lsi\ and \hj, the precision of 10-$\mu$as
level is enough for judging their compact objects.
In the case of such astrometric precision, the confidence levels for the measured mass of 1.4, 5.0, and 10 $\sm$
are 97\% or larger for \lsi\ and 99\% or larger for \hj.



The optical astrometric satellite, Gaia, is now operated
by European Space Agency, and
in the future, Small-JASMINE planned in Japanese group will also
perform astrometric measurements.
These missions aim to achieve the astrometric precision
of the 10-$\mu$as level.
Therefore, these missions could contribute the realization of
these science objectives.

\section*{Acknowledgements}
We are grateful to H. Asada, K. Yamada, and T. Hara for useful discussions
and T. Mihara for useful comments.
This work was supported by the JSPS KAKENHI Grant Number JP23244034 and JP15H02075
(both are Grant-in Aid for Scientific Research (A)).








\appendix

\section{Precision of the semi-major axis}\label{sec:sigma_a}
Here, we derive the precision of the semi-major axis, $\sigma_a$, by giving a certain
set of orbital elements $\vth \equiv (a, e, i, \omega, \Omega)$,
where $a, e, i, \omega, \text{ and } \Omega$ are the semi-major axis,
eccentricity, inclination angle, argument of periapsis, and longitude
of the ascending node, respectively.
Here, we note that the orbital period $P$ is not listed in $\vth$.
This is because orbital periods of systems focused in this paper
(nearby \hmxb s and gamma-ray binaries) have been measured with small
(typically less than a few percent) uncertainties, so that,
in this paper, we assume that the orbital period is a known element.
In what follows, we identify the precision of each orbital element
with the standard deviation $\sigma_k \ (k=1, 2,3,4 , 5)$,
where the integer $k=$ 1 corresponds to $a$,
2 to $e$, 3 to $i$, 4 to $\omega$, and 5 to $\Omega$.
The dispersion of each orbital element $\sigma^2_k$ can be represented by using
each diagonal component of a variance matrix $V$, i.e.,
\begin{equation}
\sigma^2_k = (V)_{kk}, \label{eq:sigma}
\end{equation}
The variance matrix $V$ in Equation (\ref{eq:sigma})
can be obtained by calculating variances for estimated values
of orbital elements $\vth$, so that it depends on methods for
the estimation.

If we adopt the least-square method as an estimation method
in deriving the orbital elements, then we obtain
\begin{equation}
V = (D^{\text{T}}D)^{-1} \sigma^2, \label{eq:variance}
\end{equation}
where $\sigma^2$ is a dispersion of the positional data
and a matrix $D$ represents a design matrix, whose components are
the derivatives of the elliptical orbit on the celestial plane
with respect to the orbital elements (see Appendix \ref{sec:design}).
Here, we note that the components of $D$ are two-dimensional vectors and that
the multiplication between a component of $D^{\text{T}}$ and
one of $D$ in Equation (\ref{eq:variance}) is an inner product.
The reason why we use the least-square method is that
we can use the information of both spatial and time with this method and
that the expression of the dispersion (Equation \ref{eq:variance})
is simple enough for us to analytically obtain precisions of
orbital elements.
On the other hand, in the moment approach, which is another
way to estimate the orbital elements and developed by \citet{iwa13,yam14},
the timing information is not basically taken into account,
so that we obtain the resultant orbital elements with relatively
large uncertainties.
Thus, this approach will be used to find the trial values of the
parameters at the initial stage of
an iteration process in the least-square method.
Here, we note that the equation of the orbit 
is a non-linear equation, so that we need to make use of the
steepest descent method or Gauss-Newton method, which include
the iteration process, to find a set of
estimated values of orbital elements.

Thus, the relation between the astrometric precision $\sigma$ and
the precision of the semi-major axis is represented as
\begin{equation}
\sigma^2_1 = (D^{\text{T}}D)^{-1}_{11} \sigma^2, \label{eq:sigma_a}
\end{equation}
where the subscription, 11, represents the (1,1) component
of the matrix.
A point to notice is that the set of the estimated orbital elements $\vth$,
which is an argument of the matrix $D$,
will be obtained only after we analyse actual observational data.
Therefore, in Section \ref{sec:results} we adopt
known orbital elements obtained by observations of radial velocities
for some specific binaries.

We can further reduce Equation (\ref{eq:sigma_a}) to obtain
the expression (see Appendix \ref{sec:order}):
\begin{equation}
\sigma_a = f(e,i,\omega)\frac{\sigma}{\sqrt{N}}, \label{eq:osiga}
\end{equation}
where the precision $\sigma_1$ is replaced to $\sigma_a$ and
$N$ is the number of the positional data for the optical star.
The coefficient $f$ in Equation (\ref{eq:osiga}) is
a function of the orbital elements
$e,i,$ and $\omega$ (Appendix \ref{sec:order}) and accurately calculated by
a procedure in Appendix \ref{sec:cal}.

\begin{figure}
 \begin{center}
 \includegraphics[width=8cm,clip]{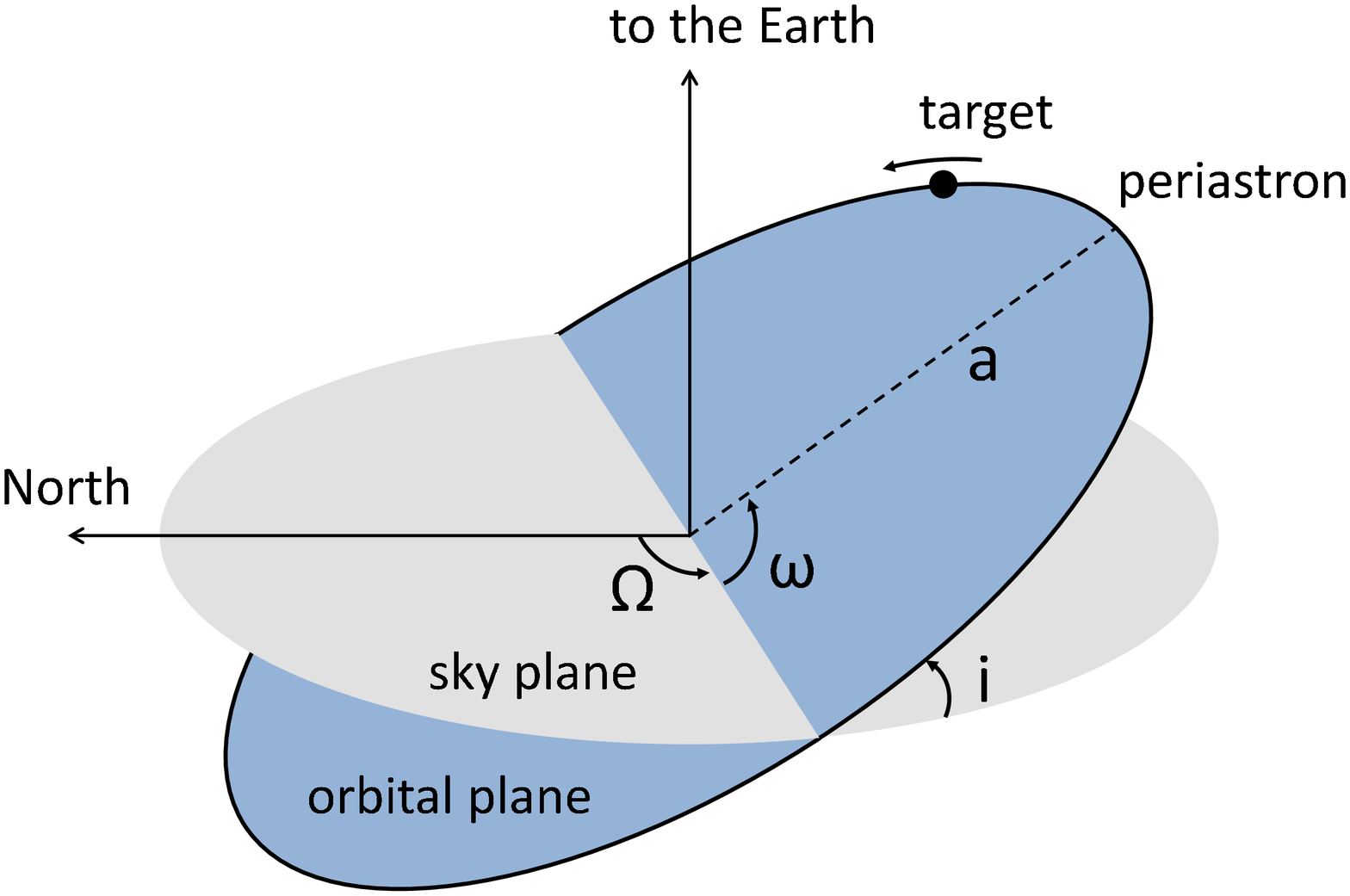}
 \caption{Sketch of an elliptical orbit of a target star in a binary.
 The gray and blue surfaces are the sky plane and the orbital plane, respectively.
 The semi-major axis is shown with a dashed line whose length is $a$,
 and its direction on the orbital plane is characterized by the argument of
 periapsis $\omega$.
 The normal direction to the orbital plane is characterized by the
 inclination angle $i$ and the longitude of ascending node $\Omega$.
  The direction ``North'' represents the direction to the celestial north pole.
 }
 \label{fig:orbit}
 \end{center}
\end{figure}

\section{Elliptical orbit on the celestial sphere}
In this section we formulate the elliptical motion on the celestial sphere.
First, we present an elliptical orbit of a target star
in a binary projected on the sky plane (the tangential plane of
the celestial sphere at the target star) $\vecr{}$;
\begin{equation}
\vecr{} = T \rell, \label{eq:vecr}
\end{equation}
where $T$ is the matrix whose components are Thiele-Innes elements
\cite[e.g.,][]{ait64};
\begin{gather}
T = \begin{pmatrix} A & F \\ B & G \end{pmatrix} \\
A = a(\co \cO - \mu \so \sO) \\
B = a(\co \sO + \mu\so \cO) \\
F = a(- \so \cO - \mu \co \sO) \\
G = a(- \so \sO + \mu \co \cO).
\end{gather}
Here $a, \mu, \omega, \Omega$ are
the angular distance of the semi-major axis, the cosine of the inclination angle
$i$, the argument of periapsis and the longitude of ascending node, respectively
(Figure \ref{fig:orbit}).
The vector $\rell$ in Equation (\ref{eq:vecr}) represents the elliptical orbit
on the orbital plane whose semi-major axis is normalised as unity and
whose focal point is at the origin, that is,
\begin{equation}
\rell = \begin{pmatrix} \cos E -e \\ \sqrt{1-e^2} \sin E \end{pmatrix},
\end{equation}
where $E$ and $e$ is the eccentric anomaly and the eccentricity, respectively.
Note that in Equation (\ref{eq:vecr}) we assume that we can
take the frame moving with the helical motion which consists of
the proper motion and the elliptical annual motion of the target star.
This assumption is valid if the observational
period is larger than one year, if the orbital period of the binary is
comparable to or shorter
than the observational period, and if the orbital shape or phase of the binary
is different from that of the annual elliptical motion.
Under these two conditions, the proper motion and parallax are measured
separately from the orbital motion.
Moreover, we note that the binary orbit is determined if five orbital
elements, $a, e, i, \omega, \Omega$ is fixed.

Given observational data, the orbital equation (Equation \ref{eq:vecr})
includes two more elements.
The time $t$ is related to the eccentric anomaly $E$ through the mean anomaly
$M$ as follows,
\begin{equation}
M = 2\pi \frac{t-t_0}{P} = E-e\sin E \label{eq:ma},
\end{equation}
where $t_0$ and $P$ is the time of periastron passage and the orbital period,
respectively.
Here, we assume that $t_0$ and $P$ are obtained
by the observation of radial velocity due to the orbital motion,
but in this paper we pick up only the binaries whose two elements, $t_0$ and $P$,
have been already measured.
Thus, $E$ is uniquely determined if the observational time is given.

\section{Design matrix for the least square method} \label{sec:design}
Here, we linearize Equation (\ref{eq:vecr}) to apply the least-square method
to estimation of the orbital elements and their dispersions.
Hereafter, we express the orbital elements as a column vector,
$\vth \equiv \begin{pmatrix} a & e & \mu & \omega & \Omega
\end{pmatrix}^{\text{T}}$, where the superscript T represents transposition.
We will obtain positions of a target star on the celestial sphere at many times
provided by Small-JASMINE and Gaia.
We define $N$ as the number of the positional data,
$\vx{j}$ as the $j$-th positional data which do not include the proper
motion and the annual elliptical motion, and $E_j$ as the $j$-th
eccentric anomaly corresponding to the time of the $j$-th positional data.
Thus, we can express $\vx{j}$ as
\begin{equation}
\vx{j} = \vecr{} (E_j; \vth) + \veps{j} \ (j=1, \dots, N),
\end{equation}
where $\veps{j}$ represents a residual vector of the $j$-th observational data.
By expanding $\vecr{} (E_j; \vth)$ around a set of trial orbital elements $\vtho$,
we find
\begin{equation}
\vx{j} = \vecr{} (E_j; \vtho) + D_j \cdot d\vth + \veps{j}\ (j=1, \dots, N),
\end{equation}
where
\begin{equation}
D_j = \pf{\vecr{}(E_j; \vth)}{\vth} \bigg| _{\vth=\vtho},
\end{equation}
and $d\vth$ represents a column vector of differentials for the orbital
elements.
We define a differential position $d\vx{j} \equiv \vx{j} - \vecr{}
(E_j; \vtho)$ and
gather $N$ equations of the positional data together, so that we obtain
\begin{equation}
dX = D d\vth + \epsilon, \label{eq:design}
\end{equation}
where
\begin{gather}
dX \equiv \left( d\vx{1} \ \dots \ d\vx{N} \right)^{\text{T}} \\
D \equiv
  \begin{pmatrix}
  \pf{\vecr{1}}{a} & \pf{\vecr{1}}{e} & \pf{\vecr{1}}{\mu} & 
  \pf{\vecr{1}}{\omega} & \pf{\vecr{1}}{\Omega}
  \\ & & \vdots & & \\
  \pf{\vecr{N}}{a} & \pf{\vecr{N}}{e} & \pf{\vecr{N}}{\mu} & 
  \pf{\vecr{N}}{\omega} & \pf{\vecr{N}}{\Omega} 
  \end{pmatrix}\label{eq:dmatrix}\\
\epsilon \equiv \left( \veps{1} \ \dots \ \veps{N} \right)^{\text{T}},
\end{gather}
where $\vecr{j}\ (j=1, \dots, N)$ in the matrix $D$ represent $\vecr{}(E_j; \vth)$.
We note that the $N \times 5$ matrix $D$, which is called a design matrix,
depends on $\vtho$ and that its components are two-dimensional vectors.
We adopt $d\vth$ at which the norm of $\epsilon$ reaches a minimum
as the estimated value.
If we assign the value $\vtho + d\vth$ to $\vtho$ in Equation (\ref{eq:design})
to find a new $d\vth$ at which the norm of $\epsilon$ reaches a minimum,
then we obtain better solution for the orbital elements.
By iterating this procedure until $d\vth$ converges at 0, we find
the estimated value of $\vth$.
In the actual analysis, we will stop the iteration when $|d\vth|/|\vtho|$ is
sufficiently smaller than unity, e.g., 0.01.

\section{Calculational procedure of precisions of orbital elements}\label{sec:cal}
Here, we describe how to compute the precisions $\sigma_k$
(Equation \ref{eq:sigma}),
taking into account the actual observation.
In Equation (\ref{eq:variance}), we have shown that the precisions
of orbital elements are represented using $D^{\text{T}}D$,
where the expression of the matrix $D$ is represented in Appendix \ref{sec:design}.
Thus, we should reduce $D^{\text{T}}D$ to obtain the precisions.

We first assume that all time intervals between two consecutive data
are equal, so that we can replace approximately the summation
in the components of the 5$\times$5 matrix $D^{\text{T}}D$
with integration with respect to time.
We can express an $lm$ component of the matrix $D^{\text{T}}D$ as
\begin{align}
(D^{\text{T}}D)_{lm}
 &=    \sum_{j=1}^N \pf{\vecr{j}}{\theta_l}   \cdot \pf{\vecr{j}}{\theta_m} 
 \notag \\
 &= \frac{N}{P}
 \int_{t_0}^{t_0+P} \pf{\vecr{}(E)}{\theta_l} \cdot \pf{\vecr{}(E)}{\theta_m}
 dt,
 \label{eq:nsum}
\end{align}
where we note that the eccentric anomaly $E$ is a single-valued function
of the time $t$. 
Equation (\ref{eq:nsum}) is valid when the orbital period is shorter than
the observational period, i.e., the time it takes to obtain N positional data.
For example, Gaia observes an identical object by $\sim$ 80 times for 5 years
\citep{deb12}, so that we can apply it to binaries with $P < 5$ yrs.

Equation (\ref{eq:nsum}) is further reduced by a variable transformation,
\begin{equation}
 dt = \frac{dt}{dE} dE = \frac{P}{2\pi}(1-e\cos E) dE.
\end{equation}
Using this relation, we obtain the expression,
\begin{equation}
(D^{\text{T}}D)_{lm}
 = \frac{N}{2\pi} 
 \int_{0}^{2\pi} \pf{\vecr{}(E)}{\theta_l} \cdot \pf{\vecr{}(E)}{\theta_m}
 (1- e\cos E) dE. \label{eq:dtd}
\end{equation}
Since the integrand in Equation (\ref{eq:dtd}) is represented with a quadratic
of sine and cosine, the integration is readily performed.
Thus, we obtain the precisions of orbital elements
by computing the inverse matrix of $D^{\text{T}}D$ using Equation
(\ref{eq:dtd}).

\section{Order estimation of the precisions}\label{sec:order}
Here, we estimate the orders of $\sigma_k \ (k=1,\dots, 5)$ before
the accurate calculations.
The order of $\left| \partial \vecr{}/\partial \theta_1 \right|$ are 1 because
it is rewritten as $ \left|\vecr{}\right|/a$, while the orders
of $\left| \partial \vecr{}/\partial \theta_k \right| \ (k=2,3,4,5)$ are $a$.
Thus, from Equation (\ref{eq:dtd})
the orders of $(D^{\text{T}}D)_{lm}$ are as follows.
\begin{equation}
(D^{\text{T}}D)_{lm} = \begin{cases}
O(1) N & \text{for } l=m=1 \\
O(1) a N & \text{for $l=1$ or $m=1$} \\
O(1) a^2 N & \text{for $l\neq 1$ and $m\neq 1$}
\end{cases}. \label{eq:orderdtd}
\end{equation}
This leads to the fact that $(D^{\text{T}}D)^{-1}_{11}$ is proportional to
$N^{-1}$, while $(D^{\text{T}}D)^{-1}_{ll}\ (l=2,3,4,5)$ are $(Na^2)^{-1}$.
Thus, we obtain the order of the precisions $\sigma_k$ as
\begin{equation}
\sigma_k = \frac{\sigma}{\sqrt{N}} \begin{cases}
O(1)  & \text{for } k=1 \\
O(1)a^{-1} & \text{for }k\neq 1
\end{cases}. \label{eq:ordersigma}
\end{equation}
This implies that the precision of the semi-major axis does not depend on
the semi-major axis itself, which is natural because the precision of the
semi-major axis is equivalent to the precision of the positional
determination by the astrometric measurements.
We also see that the precisions of the other orbital elements
are inversely proportional to the semi-major axis.
In addition, it is reasonable that the precisions are inversely proportional
to $\sqrt{N}$, because the residual of the data $\epsilon$ is assumed to
be statistical errors.
Since $\epsilon$ actually includes systematic errors, $\sigma/\sqrt{N}$
approaches asymptotically to a value dependent on the systematic errors
as $N$ increases.

Thus, the precisions in Table \ref{tab:errors} correspond to 
the values of $O(1)$ in Equation (\ref{eq:ordersigma}),
which are independent of $a$.
In addition, these precisions are independent of the orbital
element $\Omega$, because the change of $\Omega$ corresponds to the rotation
of the orbit around the line of sight,
so that it never affects the shape of the orbit on the sky plane.
The precisions of orbital elements are unchanged if the shape of the orbit
unchanged.
This independence is verified by the reduction of the integrand
in Equation (\ref{eq:dtd}).
Therefore, $\sigma_k \ (k=1,\dots, 5)$ are uniquely determined by the orbital
elements $e,i,$ and $\omega$, which enable us to present the
expression for $\sigma_k \ (k=1,\dots, 5)$:
\begin{equation}
\sigma_k = \frac{\sigma}{\sqrt{N}} \begin{cases}
f(e,i,\omega)  & \text{for } k=1 \\
g(e,i,\omega)a^{-1} & \text{for }k\neq 1
\end{cases}, \label{eq:fandg}
\end{equation}
where
\begin{equation}
\begin{split}
f &= \sqrt{N(D^{\text{T}}D)^{-1}_{11}} & \text{for } k=1 \\
g &= a\sqrt{N(D^{\text{T}}D)^{-1}_{kk}} & \text{for }k\neq 1
\end{split}.
\end{equation}

\section{Calculational results of the precision for each orbital element}
\label{sec:calres}
We show here the results obtained by calculating the precisions of
orbital elements $\sigma_k$ (Equation \ref{eq:sigma}).
We first investigate general features of the precisions.
In Table \ref{tab:errors} we show the precisions for 4 cases:
$(e,i) = (0.2,30^{\circ}), (0.2,60^{\circ}), (0.8, 30^{\circ}),$ and
$(0.8,60^{\circ})$.
We adopt a constant value 
0$^{\circ}$ for $\omega$ in Table \ref{tab:errors}.
The units of precisions are represented as $\sigma/\sqrt{N}$ for $k=1$
and as $\sigma/a\sqrt{N}$ for $k\neq 1$.
We note that the precision of the semi-major axis $\sigma_a$
corresponds to $f(e,i,\omega)$ in Equation (\ref{eq:errprop_red}).

We see from Table \ref{tab:errors} that the precision of
the semi-major axis is independent of the inclination angle.
This is because the change in the inclination angle does not
affect the angular
size of the semi-major axis when $\omega = 0^{\circ}$, which is the case
that the semi-major
axis has an overlap with the line of intersection between the sky plane and
the orbital plane. 
In addition, $\sigma_a$ is nearly independent of the eccentricity,
because the major axis is unchanged with the change of the eccentricity.
Thus, when $\omega = 0^{\circ}$, the precision of the semi-major axis is
nearly unchanged.

Table \ref{tab:errors} also shows that the precisions of $\omega$ and
$\Omega$ decrease with increase in the inclination angle.
For a smaller $i$, which corresponds to $\mu$ close to 1,
the matrix $T$ is approximately a function of $a$ and $\omega+\Omega$.
This means that the shape of the orbit on the sky plane depends on
$\omega+\Omega$, so that $\omega$ and $\Omega$ cannot be solved separately
from the shape.
Thus, the precisions of $\omega$ and $\Omega$ are enhanced.
On the other hand, for a larger $i$, the shape of orbit 
is more dependent on both $\omega$ and $\Omega$, so that their precisions
are reduced.
Here we should note that the change in
the element $\omega$ has little effect on the shape of orbit
in the case of a small $e$.
Although it seems that this means a large precision of $\omega$,
the precision $\sigma_{\omega}$ appeared in Table \ref{tab:errors} is
comparable to the other precisions when $e=0.2$ and $i=60^{\circ}$.
This is because difference of the velocity between different orbital phases
helps $\omega$ to be determined.

In Table \ref{tab:errors2}, we show the precisions when changing
the argument of periapsis.
We calculate the precisions for $\omega=90^{\circ}$, which corresponds
the situation where the major axis is perpendicular to the line of
intersection between the sky plane and the orbital plane.
Here we adopt $e=0.2 \text{ and }0.8$ with a fixed $i$ (30$^{\circ}$).
When $e=0.2$, all precisions are nearly unchanged even if $\omega$ changes,
because the shape of the orbit nearly unchanged even if $\omega$ changes
when the orbit is nearly circular.
On the other hand, when $e=0.8$, Table \ref{tab:errors2} shows that
the precision of the semi-major axis is clearly larger than that when
$\omega=0^{\circ}$.
This is because the changes in $a$ and in $i$ have a similar effect on
the shape of orbit, so that it is difficult to determine $a$ and $i$
separately.
Thus, the precision of the semi-major axis is relatively large
in the case of
the large eccentricity and the argument of periapsis near 90$^{\circ}$.

The precision of the semi-major axis depends on the inclination
angle when the eccentricity is large and the argument of periapsis
is close to 90$^{\circ}$.
In Table \ref{tab:errors3}, we show the dependence of $\sigma_a$
on $i$ and $\omega$ when $e=0.8$.
When $\omega=0$, $\sigma_a$ do not depend on the value of $i$
even the eccentricity is large and show the value of 1.7,
which is consistent with the third and fourth lines in
Table \ref{tab:errors}.
On the other hand, $\sigma_a$ depends on $i$ when $\omega$ 
is large, and the range of $\sigma_a$ increases as $\omega$
is close to 90$^{\circ}$, which are shown in the second
and third lines in Table \ref{tab:errors3}.
This is due to the size of the orbit.
When the eccentricity is large and $\omega$ is close to
90$^{\circ}$, the shape of the orbit on the celestial sphere
significantly changes as the inclination angle changes.
The orbit shows ellipse with large eccentricity when
$i\sim 0$, whereas the shape of orbit appears to approach a circle
with smaller semi-major axis than the original ellipse
as $i$ becomes large.
If the apparent size of orbit becomes smaller, it becomes harder
to measure the orbital elements by astrometry.
This leads to the fact that $\sigma_a$ increases as the inclination
angle increases, which is more significant for $\omega$ closer
to 90$^{\circ}$.
Therefore, $\sigma_a$ show the large dependence on the inclination
angle when the eccentricity is large and the argument of periapsis
is close to 90$^{\circ}$.

\begin{table}
\caption{Calculated precisions when $\omega = 0^{\circ}$.
We calculate them for 4 cases in which
eccentricity and inclination angle take two different values. The number in
subscripts of $\sigma$ are replaced with the orbital elements itself.
The precision $\sigma_a$ is normalised with $\frac{\sigma}{\sqrt{N}}$, and
those of the other orbital elements are normalised with $\frac{\sigma}
{a\sqrt{N}}$.}
\label{tab:errors}      
\centering                          
\begin{tabular}{llrrrrr}        
\hline\hline                 
  $e$ & $i$ \phantom{0}& $\sigma_a$ & $\sigma_e$ & $\sigma_{\mu}$
  & $\sigma_{\omega}$ & $\sigma_{\Omega}$  \\
\hline                        
   0.2 & 30$^{\circ}$ \phantom{0}& 1.4 & 1.1 & 2.0 & 7.4 & 7.3 \\
   0.2 & 60$^{\circ}$ \phantom{0}& 1.4 & 1.1 & 1.6 & 2.1 & 2.0 \\
   0.8 & 30$^{\circ}$ \phantom{0}& 1.7 & 2.3 & 6.2 & 9.8 & 8.9 \\
   0.8 & 60$^{\circ}$ \phantom{0}& 1.7 & 2.3 & 4.1 & 3.2 & 1.9 \\
\hline                                   
\end{tabular}
\end{table}

\begin{table}
\caption{Calculated precisions when $i = 30^{\circ}$.
The value of $a$ is the same as Table \ref{tab:errors}.
We calculate them for 4 cases in which
eccentricity and argument of periapsis take two different values.
The number in
subscripts of $\sigma$ are replaced with the orbital elements itself.
The normalisation of precisions is the same as Table \ref{tab:errors}.
The values in the first and third lines are the same as Table
\ref{tab:errors}.}
\centering                          
 \begin{tabular}{llrrrrr}
\hline\hline                 
  $e$ & $\omega$ \phantom{0}& $\sigma_a$ & $\sigma_e$ & $\sigma_{\mu}$
  & $\sigma_{\omega}$ & $\sigma_{\Omega}$ \\
\hline                        
   0.2 & \phantom{0}0$^{\circ}$ \phantom{0}& 1.4 & 1.1 & 2.0 & 7.4 & 7.3 \\
   0.2 & 90$^{\circ}$ \phantom{0}& 1.4 & 1.3 & 2.0 & 7.3 & 7.4 \\
   0.8 & \phantom{0}0$^{\circ}$ \phantom{0}& 1.7 & 2.3 & 6.2 & 9.8 & 8.9 \\
   0.8 & 90$^{\circ}$ \phantom{0}& 6.3 & 2.6 & 7.0 & 8.7 & 9.8 \\
\hline                        
 \end{tabular} \label{tab:errors2}
\end{table}

\begin{table}
\caption{Calculated precisions of the semi-major axis when $e = 0.8$.
The value of $a$ is the same as Table \ref{tab:errors}.
We calculate $\sigma_a$ for 3 cases in which the values of
the argument of periapsis are 0$^{\circ}$, 40$^{\circ}$,
and 80$^{\circ}$.
The values of 10$^{\circ}$ and 80$^{\circ}$ are adopted as
the inclination angle for each value of the argument of periapsis.
The normalisation of precisions is the same as Table \ref{tab:errors}.}
\centering                          
 \begin{tabular}{llrr}
\hline\hline                 
  $e$ & $\omega$ \phantom{0}& $\sigma_a$ ($i=10^{\circ}$)
  & $\sigma_e$ ($i=80^{\circ}$)  \\
\hline                        
   0.8 & \phantom{0}0$^{\circ}$ \phantom{0}& 1.7 & 1.7  \\
   0.8 & 40$^{\circ}$ \phantom{0}& 2.0 & 2.2  \\
   0.8 & 80$^{\circ}$ \phantom{0}& 5.5 & 20 \\
\hline                        
 \end{tabular} \label{tab:errors3}
\end{table}


\bsp	
\label{lastpage}
\end{document}